\documentclass{JHEP3}
\usepackage[dvips]{graphicx}
\usepackage{amsmath}

\newcommand{\bx}{\boldsymbol{x}}
\newcommand{\by}{\boldsymbol{y}}
\newcommand{\bz}{\boldsymbol{z}}
\newcommand{\rmi}{\mathrm{i}}
\newcommand{\rmd}{\mathrm{d}}
\newcommand{\rme}{\mathrm{e}}
\newcommand{\Nc}{N_{\rm c}}
\newcommand{\Qs}{Q_{\rm s}}
\newcommand{\SU}{\mathop{\rm SU}}
\newcommand{\half}{{\scriptstyle\frac{1}{2}}}
\newcommand{\tr}{\mathop{\rm tr}}
\newcommand{\LQCD}{\varLambda_{\text{QCD}}}
\newcommand{\tf}{T_{\text{F}}}
\newcommand{\ta}{T_{\text{A}}}

\title{Light projectile scattering off the Color Glass Condensate}
\author{Kenji Fukushima and Yoshimasa Hidaka\\
        RIKEN BNL Research Center, Brookhaven National
        Laboratory, Upton, New York 11973, USA\\
        E-mail: \email{fuku@quark.phy.bnl.gov} and
                \email{hidaka@quark.phy.bnl.gov}
}

\preprint{RBRC-667}

\abstract{
 We systematically compute the Gaussian average of Wilson lines
 inherent in the Color Glass Condensate, which provides useful
 formulae for evaluation of the scattering amplitude in the collision
 of a light projectile and a heavy target.
}

\begin{document}

%%%%%%%%%%   Introduction   %%%%%%%%%%

\section{Introduction}

  The Wilson line is a requisite to elucidate the high-energy
collision of partons in the eikonal
approximation~\cite{Nachtmann:1991ua}.  Especially in case of
scattering between a light projectile and a heavy target such as the
electron-hadron collision, proton-nucleus collision, deuteron-nucleus
collision, and also nucleus-nucleus collision in the forward or
backward rapidity region due to small-x evolution, etc, the scattering
amplitude is expressed in terms of the Wilson line representing
partons which reside in a light projectile and travel through random
color fields from a heavy 
target~\cite{Hufner:1990uv,Fujii:2002cj,Fujii:2002vh}. It is not the
individual particles inside the target but its surrounding fields that
the projectile can probe.  Such a description is analogous to the
Weizsacker-Williams approximation in which electron is viewed as
equivalent photon.  The notion of non-Abelian analogue of the
Weizsacker-Williams field has been well developed, which is called the
Color Glass Condensate (CGC)~\cite{McLerran:1993ni,Kovchegov:1997pc}
in the field of high-energy QCD.

  The McLerran-Venugopalan (MV) model assumes that the color charge
density $\rho$ is static ($x^+$-independent), is a function of the
transverse $\bx_\perp$ and longitudinal $x^-$ coordinates, and
distributes randomly at each spatial point.  Its magnitude squared
$|\rho|^2$ should be proportional to the transverse 
density of partons consisting of a heavy target, which is commonly
denoted by $\mu^2$ in the model.  One can compute the scattering
amplitude by taking the Gaussian average of Wilson lines embodying the
projectile given a certain $\mu$ relevant to the experimental
condition.  Since the explicit form of the non-Abelian
Weizsacker-Williams field is known~\cite{Kovchegov:1997pc}, the above
mentioned is a doable calculation.

  In fact, one can find evaluation of the Gaussian average of Wilson
lines in literatures~\cite{Fujii:2002cj,Fujii:2002vh,Kovner:2001vi,%
Gelis:2001da,Blaizot:2004wv,Baier:2005dv} in different contexts and
thus with different color structure, representation, etc.  We here aim
to derive more general formulae, which provides us with useful
implements to describe high-energy collisions.  The most general form
is, as easily anticipated, too complicated to handle directly once the
number of Wilson lines is more than four, as we will encounter later
in this paper.  In that case we will attempt to simplify the
expression under the limit of large $\Nc$ where $\Nc$ is the number of
colors.  We will see that a picture of the color dipoles instead of
gluons naturally arises in the large-$\Nc$ limit.

  Just for clarity of what we will address, we prefer to use the
terminology, ``scattering amplitude'' to signify the Wilson line
correlator.  That quantity is, however, not limited only to the
scattering process but would appear in the process of particle
production from the CGC
background~\cite{Blaizot:2004wv,Baier:2005dv,Gelis:2001da,Fukushima:2007}.
Also, we would mention that the Gaussian average is not only limited
to the MV model but is widely relevant to the CGC formalism with a
Gaussian approximation~\cite{Iancu:2002aq}.  Therefore, we believe
that the potential application of our formalism should be ubiquitous
in high-energy QCD.

%%%%%%%%%%   Gaussian average of Wilson lines   %%%%%%%%%%

\section{Gaussian average of Wilson lines}

  Our goal is to derive the general expression of the Gaussian average
or correlation function in terms of Wilson lines under random
distribution of color source.  In a physical terminology the
correlation function represents the scattering amplitude of a bunch of
particles and antiparticles traveling through random color source in
the eikonal approximation.  That is, the specific quantity of our
interest in this paper is
\begin{equation}
 \bigl\langle U(\bx_{1\perp})_{\beta_1\alpha_1}
  U(\bx_{2\perp})_{\beta_2\alpha_2} \cdots
  U(\bx_{n\perp})_{\beta_n\alpha_n} \bigr\rangle \,,
\label{eq:correlator}
\end{equation}
where the Greek indices are with respect to color in a certain
representation $r$ of the $\SU(\Nc)$ group.   In the MV model the
Wilson line is written by the non-Abelian Weizsacker-Williams field
given as a solution of the classical Yang-Mills equation of motion;
\begin{equation}
 U(\bx_\perp) = \mathcal{P}\exp\Biggl[ -\rmi g^2
  \int_{-\infty}^{+\infty}\!\rmd x^-\rmd^2 \bz_\perp \,
  G_0(\bx_\perp \!-\! \bz_\perp)\, \rho_a(x^-,\bz_\perp)\, t^a
  \Biggr] \,,
\label{eq:Wilson}
\end{equation}
where $t^a$'s are color matrices of the $\SU(\Nc)$ algebra in the $r$
representation.  We denote the time ordering operator in the $x^-$
direction by $\mathcal{P}$ and the two-dimensional propagator by
$G_0(\bx_\perp)$ which satisfies the Poisson equation,
\begin{equation}
 \frac{\partial^2}{\partial\bx_\perp^2} G_0(\bx_\perp)
  = \delta^{(2)}(\bx_\perp) \,. 
\end{equation}
Figure~\ref{fig:correlator} is the schematic picture of the average
(\ref{eq:correlator}) with color indices.  The blob part is the target
which provides the random and dense $\rho_a(x^-,\bx_\perp)$ from the
target, where $x^-$ and $\bx_\perp$ indicate the spatial point on the
transverse (impact-parameter) plane and the longitudinal extent of the
target respectively.  It should be noted that $x^-$ is regarded as a
time variable for the projectile.  Thus, the Wilson line
(\ref{eq:Wilson}) encodes projectile's multiple scattering off the CGC
along the temporal $x^-$ direction.

\FIGURE[t]{
$ \bigl\langle U(\bx_{1\perp})_{\beta_1\alpha_1}
  U(\bx_{2\perp})_{\beta_2\alpha_2} \cdots
  U(\bx_{n\perp})_{\beta_n\alpha_n} \bigr\rangle
 = $~~\parbox{3.5cm}{\includegraphics[width=3.5cm]{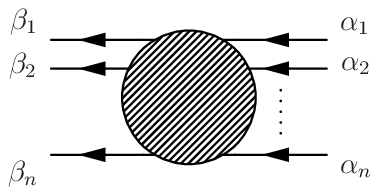}}
\caption{Graphical representation of the correlation function in
  terms of Wilson lines which corresponds to the scattering amplitude
  of $n$-particles traveling through a heavy target with the random
  and dense color distribution.  Each Wilson line stands for a parton
  which starts with the color orientation $\alpha_i$ before scattering
  and ends up with $\beta_i$ with a fixed transverse position
  $\bx_{i\perp}$ due to the eikonal approximation.}
\label{fig:correlator}
}

  It is assumed in the MV model that the average
$\langle\cdots\rangle$ is accompanied by the Gaussian weight in terms
of $\rho_a(x^-,\bx_\perp)$, whose dispersion specifies the typical
model scale $\mu$ in the standard convention, or in other words, the
saturation scale $\Qs$ related to $\mu$ up to a logarithmic factor
(see Eq.~(\ref{eq:Qs}) for our definition without logarithm)
universally characterizes the hadron wavefunction.  As we mentioned
before, we will develop our method for the MV model for example, but
the technique is applicable to any CGC calculation with a Gaussian
weight function as adopted in Ref.~\cite{Iancu:2002aq}.

  The explicit form of the Gaussian weight is
\begin{equation}
 \omega(\rho) = \exp\Biggl[ -\int_{-\infty}^{+\infty}\!\rmd x^-
  \rmd\bx_\perp \,\frac{\rho_a^2(x^-,\bx_\perp)}{2\mu^2(x^-)} \Biggr]\,.
\label{eq:weight}
\end{equation}
In fact, the random walk in $\SU(\Nc)$ group space leads to the
quadratic term~\cite{Jeon:2004rk} in the weight function, and besides,
the cubic term~\cite{Jeon:2005cf} which is sensitive to Odderon
exchange but is beyond our current scope.

  The only necessary ingredient for our calculation in what follows
is, as a matter of fact, the two-point function of $\rho_a$ which is
spatially uncorrelated as
\begin{equation}
 \bigl\langle\rho_a(x^-,\bx_\perp)\rho_b(y^-,\by_\perp)\bigr\rangle
 =\delta_{ab}\, \delta(x^- - y^-)\, \delta^{(2)}(\bx_\perp - \by_\perp)\,
  \mu^2(x^-) \,,
\label{eq:correlator_rho}
\end{equation}
which contains the equivalent information as the weight
function~(\ref{eq:weight}).

  Now we have finished the setup of the MV model, that is, we have
explained the notation and the model definition in a self-contained
manner.  In the subsequent discussions we will proceed toward the
general expression in the $n$-particle case step-by-step starting with
the simplest case of one particle.

%%%   One-point function   %%%

\subsection{One-point function; $\bigl\langle U(\bx_\perp)_{\beta\alpha}\bigr\rangle$}

  We aim to make clear our notation (which is the same as
Ref.~\cite{Gelis:2001da}) first in a warming-up exercise though the
average of one-point function is not physically relevant.  Our
treatment and convention are parallel to Ref.~\cite{Gelis:2001da}.
Here we introduce the Wilson line integrated over a finite range
defined by
\begin{equation}
 U(b^-,a^-|\bx_\perp) = \mathcal{P}\exp\Biggl[ -\rmi g^2\int_{a^-}^{b^-}
  \!\rmd z^- \rmd^2\bz_\perp \, G_0 (\bx_\perp \!-\! \bz_\perp)\,
  \rho_a(z^-,\bz_\perp)\, t^a \Biggr] \,.
\end{equation}
The limit of $a^-\to-\infty$ and $b^-\to+\infty$ renders the above the
Wilson line as defined in Eq.~(\ref{eq:Wilson}).  We will expand the
finite ranged Wilson line and compute its Gaussian average using
Eq.~(\ref{eq:correlator_rho}).  The Taylor expansion of time-ordered
exponential function leads to
\begin{eqnarray}
 \bigl\langle U(b^-,a^-|\bx_\perp) \bigr\rangle &=&
  \sum_{n=0}^\infty (-\rmi g^2)^n\int\prod_{i=1}^n
  \rmd^2\bz_{i\perp}\, G_0(\bx_\perp \!-\! \bz_{i\perp})
  \int^{b^-}_{a^-}\!\rmd z_1^- \int^{z^-_1}_{a^-}\!\rmd z_2^-
  \cdots \int^{z_{n-1}^-}_{a^-}\!\rmd z_n^- \times\nonumber\\
 &&\times \bigl\langle\rho_{a_1}(z_1^-,\bz_{1\perp})
  \rho_{a_2}(z_2^-,\bz_{2\perp})\cdots \rho_{a_n}(z_n^-,\bz_{n\perp})
  \bigr\rangle\, t^{a_1}\,t^{a_2}\cdots t^{a_n} \,. 
\label{eq:averageU}
\end{eqnarray}
Here we can decompose $\langle\cdots \rangle$ into all possible
contractions in case of the Gaussian average.  Then, only the adjacent
contraction making the \emph{tadpole} diagram as shown in
Fig.~\ref{fig:diagram}~(a) survives and other contractions as in
Figs.~\ref{fig:diagram}~(b) and (c) vanish because of the
delta-function in Eq.~(\ref{eq:correlator_rho}).  Since the tadpole
contribution is to be factorized as
$\langle\rho_{a_1}\rho_{a_2}\cdots\rho_{a_n}\rangle=\langle%
\rho_{a_1}\rho_{a_2}\rangle\langle\rho_{a_3}\cdots\rho_{a_n}\rangle$,
we can rewrite Eq.~(\ref{eq:averageU}) into a form of the integral
equation;
\begin{eqnarray}
&& \hspace{-1cm}
\bigl\langle U(b^-,a^-|\bx_\perp)_{\beta\alpha} \bigr\rangle
  = \nonumber\\
&=& \delta_{\beta\alpha} + (-\rmi g^2 )^2 \int_{a^-}^{b^-}\!\rmd z_1^-
   \rmd^2\bz_{1\perp} \int_{a^-}^{z_1^-}\!\rmd z_2^- \rmd^2\bz_{2\perp}\,
   G_0(\bx_\perp \!-\! \bz_{1\perp})
   G_0(\bx_\perp \!-\! \bz_{2\perp})\times \nonumber\\
&&
 \times \bigl\langle \rho_{a_1}(z_1^-,\bz_{1\perp})
  \rho_{a_2}(z_2^-,\bz_{2\perp})\bigr\rangle
  \bigl(t^{a_1} t^{a_2}\bigr)_{\beta\gamma}
  \bigl\langle U(z_2^-,a^-|\bx_\perp)_{\gamma\alpha}
  \bigr\rangle  \nonumber\\
&=& \delta_{\beta\alpha}-\frac{g^4}{2} C_2(r)\,
  \delta_{\beta\gamma} \int\rmd^2\bz_\perp\,
  G_0^2(\bx_\perp \!-\! \bz_\perp)
  \int^{b^-}_{a^-}\!\rmd z^- \mu^2(z^-) \bigl\langle
  U(z^-,a^-|\bx_\perp)_{\gamma\alpha} \bigr\rangle \,,
\label{eq:onepoint}
\end{eqnarray}
where we note that we used
$\int^{z_1^-}_{a^-}\!\rmd z_2^-\,\delta(z_1^-\!-\!z_2^-)=\frac{1}{2}$
and $(t^a t^a)_{\beta\alpha}=C_2(r)\delta_{\beta\alpha}$ with the
second-order Casimir invariant $C_2(r)$ in the $r$ representation.
\FIGURE[t]{
 \begin{tabular}{cp{5mm}cp{5mm}c}
 \includegraphics[width=3.cm]{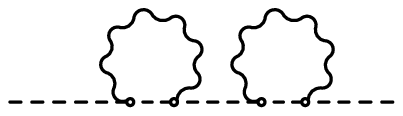} &&
 \includegraphics[width=3.5cm]{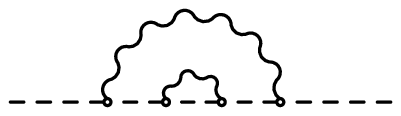} &&
 \includegraphics[width=3.5cm]{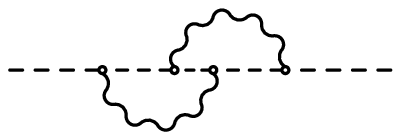} \\
 (a) &&  (b) && (c)
 \end{tabular}
\caption{Contraction of four sources; (a) tadpole type, (b) nesting
 one, and (c) overlapping one.  Two dots connected by the wavy line
 are contracted to the same point (time) and only tadpole-type
 diagrams remain finite.}
\label{fig:diagram}
}
We show the diagrammatic representation of this integral equation as
\FIGURE[h]{
 \includegraphics[width=12.5cm]{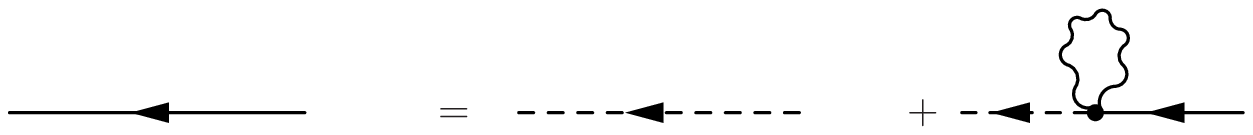}
}

\noindent
with the tadpole attached at $(z^-,\bx_\perp)$.  We can easily find
the solution, that is given by
\begin{equation}
 \bigl\langle U(b^-,a^-|\bx_\perp)_{\beta\alpha} \bigr\rangle =
  \bar{U}(b^-,a^-|\bx_\perp)\,\delta_{\beta\alpha} =
  \exp\biggl[-\Qs^2(b^-,a^-)\frac{2\Nc}{\Nc^2-1}\, C_2(r)\, L(x,x)
  \biggr]\,\delta_{\beta\alpha} \,,
\label{eq:one-point}
\end{equation}
where we defined
\begin{eqnarray}
 L(x,y) &=& g^4\int\rmd^2\bz_\perp\,
  G_0(\bx_\perp\!-\!\bz_\perp)\, G_0(\by_\perp\!-\!\bz_\perp)\,,\\
 \Qs^2(b^-,a^-) &=& \frac{\Nc^2-1}{4\Nc}\int^{b^-}_{a^-}\!\rmd z^-
  \,\mu^2(z^-) \,.
\label{eq:Qs}
\end{eqnarray}
Here we remark that we will simply write $\Qs^2$ to denote
$\Qs^2(+\infty,-\infty)$ in later discussions.
It should be mentioned that $L(x,x)$ does not depend on $x$ in fact
because of translational invariance, and thus we can write it as
$L(0,0)$ equivalently.  As we will argue later, however, $L(x,y)$
generally suffers infrared singularity, and the expectation value
(\ref{eq:one-point}) turns out to be negligible small.  This
observation intuitively corresponds to the fact that a single quark or
gluon with non-trivial color charge would interact with color charge
fluctuations inside the target at even far distance on the transverse
plane.  As a result of this long-ranged interaction (which should be
cut off by either the target size or confining scale $\sim\LQCD^{-1}$),
a quark or gluon is absorbed strongly in multiple scattering.

%%%   Two-point function   %%%

\subsection{Two-point function;
$\bigl\langle U(\bx_{1\perp})_{\beta_1\alpha_1}U(\bx_{2\perp})_{\beta_2\alpha_2}\bigr\rangle$}

\FIGURE[t]{
 \begin{tabular}{cp{4cm}c}
 \includegraphics[width=3cm]{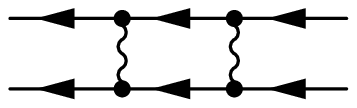} &&
 \includegraphics[width=3cm]{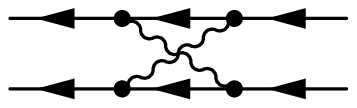} \\
 (a) && (b)
 \end{tabular}
 \caption{Contractions between different Wilson lines; (a) ladder type
 and (b) crossing type.}
\label{fig:crossed}
}

\FIGURE[t]{
 \includegraphics[width=9cm]{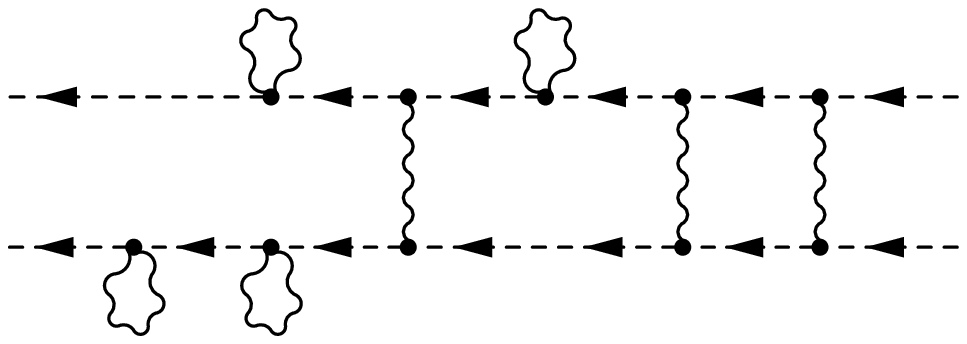}
 \caption{Typical diagram contributing to the Gaussian average of two
 Wilson lines.}
 \label{fig:typical}
}

  We shall next consider the two-point function of the Wilson lines.
This is, in contrast to the one-point function, physically relevant if
the projectile is a $q\bar{q}$ mesonic or $gg$ glueball-like state.
Not only the tadpole diagrams consisting of Fig.~\ref{fig:diagram}~(a)
but also the ladder diagrams structured with the sub-diagram
Fig.~\ref{fig:crossed}~(a) contribute to the Gaussian average (see
Fig.~\ref{fig:typical} for typical example).  Any diagram containing
the crossing sub-diagram as shown in Fig.~\ref{fig:crossed}~(b)
vanishes in the same way as the nesting and overlapping ones displayed
in Fig.~\ref{fig:diagram}.  We can compute the Gaussian average of two
Wilson lines diagrammatically by combining all the tadpole and ladder
subparts up.  The Dyson equation we need to solve is as follows;
\FIGURE[h]{
 \includegraphics[width=12cm]{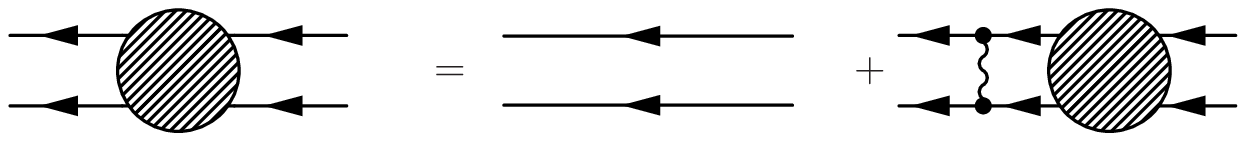}
}

\noindent
The horizontal lines represent $\langle U(b^-,a^-|\bx_\perp)\rangle$
involving all the tadpole contributions.  We can literally convert the
above graphical equation into the algebraic equation as
\begin{eqnarray}
&& \hspace{-1cm}
\bigl\langle U(b^-,a^-|\bx_{1\perp})_{\beta_1\alpha_1}
 U(b^-,a^-|\bx_{2\perp})_{\beta_2\alpha_2} \bigr\rangle =\nonumber\\
&=& \bigl\langle U(b^-,a^-|\bx_{1\perp})\bigr\rangle_{\beta_1\alpha_1}
 \bigl\langle U(b^-,a^-|\bx_{2\perp})\bigr\rangle_{\beta_2\alpha_2}
 -\nonumber\\
&& -g^4\int^{b^-}_{a^-}\!\rmd z^-\, \bigl\langle
 U(b^-,z^-|\bx_{1\perp})\bigr\rangle_{\beta_1\lambda_1} \bigl\langle
 U(b^-,z^-|\bx_{2\perp})\bigr\rangle_{\beta_2\lambda_2} \,
 t^a_{\lambda_1\gamma_1}\, t^a_{\lambda_2\gamma_2}\,
 \mu^2(z^-) \times\nonumber\\
&&\times \int\rmd\bz_\perp^2 G_0(\bx_{1\perp}\!-\!\bz_\perp)
 G_0(\bx_{2\perp}\!-\!\bz_\perp) \bigl\langle
 U(z^-,a^-|\bx_{1\perp})_{\gamma_1\alpha_1}
 U(z^-,a^-|\bx_{2\perp})_{\gamma_2\alpha_2} \bigr\rangle \nonumber\\
&=& \bar{U}(b^-,a^-|\bx_{1\perp})\, \bar{U}(b^-,a^-|\bx_{2\perp})
 \Biggl(\delta_{\beta_1\alpha_1} \delta_{\beta_2\alpha_2}
 -L(x_1,x_2) \times\nonumber\\
&& \qquad \times\int^{b^-}_{a^-}\!\rmd z^- \mu^2(z^-)\,
 t^a_{\beta_1\gamma_1}\, t^a_{\beta_2\gamma_2}
 \frac{\langle 
  U(z^-,a^-|\bx_{1\perp})_{\gamma_1\alpha_1}
  U(z^-,a^-|\bx_{2\perp})_{\gamma_2\alpha_2} \rangle}
  {\bar{U}(z^-,a^-|\bx_{1\perp})\,\bar{U}(z^-,a^-|\bx_{2\perp})}
  \Biggr) \,.
\label{eq:twopintfunc}
\end{eqnarray}
We divide the both sides of this integral equation by
$\bar{U}(b^-,a^-|\bx_{1\perp})\,\bar{U}(b^-,a^-|\bx_{2\perp})$ to
reach
\begin{eqnarray}
&& \hspace{-1cm}
F(\bx_{1\perp},\bx_{2\perp}|b^-,a^-)_{\beta_1\beta_2;\alpha_1\alpha_2}
 = \nonumber\\
&& \hspace{-1cm}
=\delta_{\beta_1\alpha_1}\delta_{\beta_2\alpha_2}
  -L(x_1,x_2)\int^{b^-}_{a^-}\!\rmd z^- \mu^2(z^-)\,
  t^a_{\beta_1\gamma_1}\,t^a_{\beta_2\gamma_2} 
  F(\bx_{1\perp},\bx_{2\perp}|z^-,a^-)_{\gamma_1\gamma_2;\alpha_1\alpha_2}
  \,,
\label{eq:F}
\end{eqnarray}
where we defined
\begin{equation}
 F(\bx_{1\perp},\bx_{2\perp}|b^-,a^-)_{\beta_1\beta_2;\alpha_1\alpha_2}
 =\frac{\langle
 U(b^-,a^-|\bx_{1\perp})_{\beta_1\alpha_1}\,
 U(b^-,a^-|\bx_{2\perp})_{\beta_2\alpha_2} \rangle}
 {\bar{U}(b^-,a^-|\bx_{1\perp})\,\bar{U}(b^-,a^-|\bx_{2\perp})} \,.
\end{equation}
The integral equation~(\ref{eq:F}) has an identical structure as
Eq.~(\ref{eq:onepoint}), so that we can solve it in the same way to
find
\begin{equation}
 F(\bx_{1\perp},\bx_{2\perp}|b^-,a^-)_{\beta_1\beta_2;\alpha_1\alpha_2}
 =\exp\biggl[ -2\Qs^2(b^-,a^-)\frac{2\Nc}{\Nc^2-1}t^a_1\,t^a_2\,
  L(x_1,x_2) \biggr]_{\beta_1\beta_2;\alpha_1\alpha_2} \,.
\end{equation}
In the limit of $a^-\to-\infty$ and $b^-\to+\infty$, as a result, we
can express the Wilson line average in the following form;
\begin{eqnarray}
 && \hspace{-1cm} \bigl\langle U(\bx_{1\perp})_{\beta_1\alpha_1}
  U(\bx_{2\perp})_{\beta_2\alpha_2} \bigr\rangle = \nonumber\\
 &=& \bar{U}(\bx_{1\perp}) \bar{U}(\bx_{2\perp})
  \exp\biggl[ -2\Qs^2 \frac{2\Nc}{\Nc^2-1}t^a_1\,t^a_2\,L(x_1,x_2)
  \biggr]_{\beta_1\beta_2;\alpha_1\alpha_2} \nonumber\\
 &=& \exp\biggl[ -\Qs^2\frac{2\Nc}{\Nc^2-1} \Bigl(2t^a_1\,t^a_2\,
  L(x_1,x_2)+ t^{a2}_1\,L(x_1,x_1)+t^{a2}_2\,L(x_2,x_2)
  \Bigr)\biggr]_{\beta_1\beta_2;\alpha_1\alpha_2} \nonumber\\
 &=& \exp\biggl[ -\Qs^2\frac{2\Nc}{\Nc^2-1} \Bigl(
  (t^a_1+t^a_2)^2 L(0,0)-t^a_1\,t^a_2\varGamma(x_1,x_2)\Bigr)
  \biggr]_{\beta_1\beta_2;\alpha_1\alpha_2} \,,
\end{eqnarray}
where we made use of translational invariance to make a shift
$L(x,x)\to L(0,0)$ and  defined
$\varGamma(x_1,x_2)=2(L(0,0)-L(x_1,x_2))$ which is free from infrared
singularity.

  Here, let us introduce ``Hamiltonian'' by
\begin{equation}
 \bigl\langle U(b^-,a^-|\bx_{1\perp})_{\beta_1\alpha_1}
  U(b^-,a^-|\bx_{2\perp})_{\beta_2\alpha_2} \bigr\rangle
 =\exp[-(H_0+V)]_{\beta_1\beta_2;\alpha_1\alpha_2}
\end{equation}
with the ``free'' part,
\begin{equation}
 H_0= \Qs^2\frac{2\Nc}{\Nc^2-1}(t^a_1+t^a_2)^2 L(0,0) \,,
\label{eq:free}
\end{equation}
and the ``interaction'' part,
\begin{equation}
 V = -\Qs^2\frac{2\Nc}{\Nc^2-1}\,t^a_1\,t^a_2 \varGamma(x_1,x_2) \,.
\label{eq:interaction}
\end{equation}
One can readily prove that $[H_0,V]=0$ meaning that $T H_0 T^{-1}$ and
$T V T^{-1}$ are to become diagonal simultaneously.  Let us consider
the decomposition of $H_0$ into irreducible representations in color
space.  We see that $H_0$ is proportional to the second-order Casimir
operator $(t^a_1+t^a_2)^2$.  In general the irreducible representation
is labeled by a set of non-negative integers $m$ with rank $\Nc-1$,
i.e., Dynkin coefficients.  The associated second-order Casimir
invariant is expressed as
\begin{equation}
 C_2(m) = \frac{1}{2\Nc}\Biggl[ \sum_{n=1}^{\Nc-1} n(\Nc-n)
  (\Nc+m_n)m_n
  + 2\sum_{n > l}^{\Nc-1} l(\Nc-n)m_n\,m_l \Biggr] \,.
\label{eq:casimir}
\end{equation}
For example $C_2=4/3$ for the fundamental representation (triplet) of
$\SU(\Nc=3)$ characterized by $m=[1,0]$ and $C_2=3$ for the adjoint
representation (octet) characterized by $m=[1,1]$.  It is obvious from
Eq.~(\ref{eq:casimir}) that $C_2(m)$ is semi-positive and zero only
when $m=0$, that is, a singlet.  If the color structure of the Wilson
line correlator is projected onto non-singlet states, $H_0$ gives a
large suppression factor, which can be seen from
\begin{eqnarray}
 L(x,y) &=& g^4\int \rmd^2\bz_\perp\, G_0(\bx_\perp \!-\! \bz_\perp)\,
  G_0(\by_\perp \!-\! \bz_\perp) \nonumber\\
 &=& g^4\int \rmd^2\bz_\perp \int\frac{\rmd^2\boldsymbol{k}_\perp}
  {(2\pi)^2} \;\rme^{\rmi(\bx_\perp-\bz_\perp)\cdot\boldsymbol{k}_\perp}
  \int\frac{\rmd^2\boldsymbol{q}_\perp}{(2\pi)^2} \;
  \rme^{\rmi(\by_\perp-\bz_\perp)\cdot\boldsymbol{q}_\perp}
  \frac{1}{\boldsymbol{k}_\perp^2}\frac{1}{\boldsymbol{q}_\perp^2}
  \nonumber\\
 &=& g^4\int\frac{\rmd k\,\rmd\theta}{(2\pi)^2}\;
  \rme^{\rmi|\by_\perp-\bx_\perp|k\cos\theta}\frac{1}{k^3}
  \nonumber\\
 &=& g^4\int_0^\infty \frac{\rmd k}{2\pi}\,\frac{1}{k^3}\,
  J_0(k|\bx_\perp-\by_\perp|) \,,
\end{eqnarray}
where $J_0(x)$ is the first-kind Bessel function.  When $k$ is small
(strictly speaking, when $k|\bx_\perp-\by_\perp|$ is small),
$J_0(k|\bx_\perp-\by_\perp|)\simeq 1-\frac{1}{4}(k|\bx_\perp-\by_\perp|)^2$,
hence, the momentum integration in $L(x,y)$ infraredly diverges.  We
introduce an infrared cutoff $\LQCD$ to regularize the infrared
singularity.  Then, we have $L(x,y)\sim 1/\LQCD^2$ and
$\varGamma(x,y)\sim|\bx_\perp\!-\!\by_\perp|^2
\ln(|\bx_\perp\!-\!\by_\perp|\LQCD)$.  Therefore, for small $\LQCD$,
$H_0$ which is proportional to $L(0,0)$ should be much larger than $V$
which is proportional to $\varGamma(x,y)$.  The non-singlet part in
the color decomposition is thus accompanied by a large suppression
factor, $\exp[-(2\Nc/(\Nc^2\!-\!1))C_2(m)\Qs^2/(4\pi\LQCD^2)]$.  This
is a physically reasonable result;  the Wilson lines form a color
singlet, and such neutral objects are free from long-ranged color
interactions (except for logarithmic singularity) which should be cut
off by the confining scale $\LQCD^{-1}$.  Because $\Qs$ (which is
$\sim\mbox{GeV}$ order) is typically greater than
$\LQCD\sim\mbox{fm}^{-1}$ by one order of magnitude at least, we do
not have to concern non-singlet parts.  From now on, accordingly, we
consider only the singlet part of the color structure of the Wilson
line product.

%%%   n-point function   %%%

\subsection{$n$-point function;
$\bigl\langle U(\bx_{1\perp})_{\beta_1\alpha_1}U(\bx_{2\perp})_{\beta_2\alpha_2}\cdots
U(\bx_{n\perp})_{\beta_n\alpha_n}\bigr\rangle$}

  Now that we have understood the computational procedure, it is easy
to generalize the formulae to the $n$-point case.  The integral
equation can be diagrammatically represented as
\FIGURE[h]{
\parbox{2cm}{\includegraphics[width=2cm]{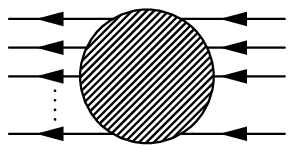}}
=
\parbox{1.8cm}{\includegraphics[width=1.8cm]{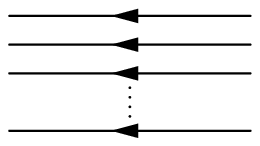}}
+
\parbox{2.6cm}{\includegraphics[width=2.6cm]{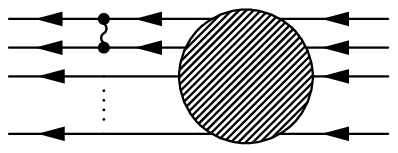}}
+
\parbox{2.6cm}{\includegraphics[width=2.6cm]{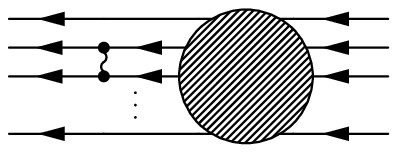}}
+
\parbox{2.6cm}{\includegraphics[width=2.6cm]{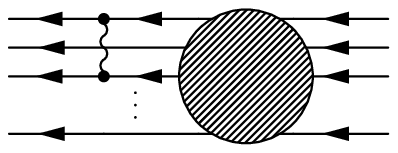}}
+$\cdots$
}

\noindent
The R.H.S.\ consists of all the possible permutations of one ladder
bridging over two out of $n$ Wilson lines.  The horizontal lines are
$\langle U(b^-,a^-|\bx_\perp)\rangle$ with all the tadpole insertions.
The corresponding equation is
\begin{eqnarray}
&& \hspace{-1cm} \bigl\langle
 U(b^-,a^-|\bx_{1\perp})_{\beta_1\alpha_1}
 U(b^-,a^-|\bx_{2\perp})_{\beta_2\alpha_2} \cdots
 U(b^-,a^-|\bx_{n\perp})_{\beta_n\alpha_n} \bigr\rangle \nonumber\\
&=& \prod_{i=1}^n \bigl\langle
 U(b^-,a^-|\bx_{i\perp}) \bigr\rangle_{\beta_i\alpha_i}
  -g^4 \sum_{i>j} \int^{b^-}_{a^-}\!\rmd z^- \mu^2(z^-)\int
  \rmd\bz_{\perp}^2\, G_0(\bx_{i\perp} \!-\! \bz_{\perp})\,
  G_0(\bx_{j\perp} \!-\! \bz_{\perp}) \times\nonumber\\
&& \times\prod_{k=1}^n \bigl\langle
 U(b^-,z^-|\bx_{i\perp}) \bigr\rangle_{\beta_k\lambda_k}\,
 \delta_{\lambda_1\gamma_1}\cdots \delta_{\lambda_{j-1}\gamma_{j-1}}\,
 t^a_{\lambda_j\gamma_j}\, \delta_{\lambda_{j+1}\gamma_{j+1}}\cdots
 \nonumber\\
&& \cdots \delta_{\lambda_{i-1}\gamma_{i-1}}\,t^a_{\lambda_i\gamma_i}
 \delta_{\lambda_{i+1}\gamma_{i+1}}\cdots
 \delta_{\lambda_n\gamma_n} \times\nonumber\\
&& \times \bigl\langle
 U(z^-,a^-|\bx_{1\perp})_{\gamma_1\alpha_1}
 U(z^-,a^-|\bx_{2\perp})_{\gamma_2\alpha_2}\cdots
 U(a^-,z^-|\bx_{n\perp})_{\gamma_n\alpha_n} \bigr\rangle \,.
\end{eqnarray}
The above integral equation takes a similar form to the case of
two-point function in Eq.~(\ref{eq:twopintfunc}), so that we can find
the solution in the same way as
\begin{equation}
 \bigl\langle U(\bx_{1\perp})_{\beta_1\alpha_1}
  U(\bx_{2\perp})_{\beta_2\alpha_2}\cdots
  U(\bx_{n\perp})_{\beta_n\alpha_n} \bigr\rangle
  =\exp[-(H_0+V)]_{\beta_1\cdots\beta_n;\alpha_1\cdots\alpha_n}
\end{equation}
with
\begin{eqnarray}
 H_0 &=& \Qs^2\frac{2\Nc}{\Nc^2-1}\biggl(\sum_{k=1}^n t_k^a\biggr)^2
  L(0,0) \,,\\
 V &=& -\Qs^2\frac{2\Nc}{\Nc^2-1}\sum_{i>j}\, t^a_i\,t^a_j\;
  \varGamma(\bx_{i\perp},\bx_{j\perp}) \,.
\end{eqnarray}
These expressions are plain generalization of
Eqs.~(\ref{eq:free}) and (\ref{eq:interaction}).  Again, $H_0$ is
proportional to the second-order Casimir operator, and after
decomposing the color structure into irreducible representations, we
can drop non-singlet parts.  What we should do to simplify the result
further is find the singlets out of the direct product of $\SU(\Nc)$
matrices which make $H_0$ vanishing.  For example, in case of the
four-point function of Wilson lines in the adjoint representation with
$\Nc=3$ (i.e.\ four gluon propagation through a dense target), there
are eight independent singlets out of $8\otimes 8\otimes 8\otimes 8$.
Thus, the Wilson line correlator can be expressed to be an $8\times8$
matrix of $\rme^{-V}$ in the basis of eight singlets.  We will face
with concrete calculations later.  In most cases of our interest in
physics problems, all we need to know is expressed in terms of its
eigenvalues and eigenstates.

%%%%%%%%%%   Examples   %%%%%%%%%%

\section{Examples}

  We will elaborate several concrete examples relevant to physical
processes.  We will identify the singlet basis to compute the
eigenvalue of the color matrix.

%%%   <UU^*> in the fundamental representation   %%%

\subsection{$\langle U(\bx_{1\perp})_{\beta_1\alpha_1}
 U^\ast(\bx_{2\perp})_{\beta_2\alpha_2} \rangle$ in the fundamental
 representation}

  In our formulae~(\ref{eq:free}) and (\ref{eq:interaction}) we set
$t_1^a=\tf^a$ and $t^a_2=-\tf^{a\ast}$ where $\tf^a$'s are the
  $\SU(\Nc)$ generator in the fundamental representation.  This
expectation value appears in one color dipole or $q\bar{q}$ scattering
off a dense target.  In this case the number of singlet is only one;
$\Nc\otimes\Nc^\ast = 1\oplus\Nc^2-1$.  If we denote the singlet state
as $|s\rangle$ then we have
$\langle\alpha_1\alpha_2|s\rangle=\delta_{\alpha_1\alpha_2}/\sqrt{\Nc}$
with a proper normalization.  The projected element of the necessary
part is
\begin{equation}
 \langle s|V|s\rangle = \Qs^2\frac{2\Nc}{\Nc^2-1}\,
  {\tf}^a_{\beta_1\alpha_1}\,{\tf}^{\ast a}_{\beta_2\alpha_2}\,
  \varGamma(x_1,x_2)\,
  \frac{\delta_{\alpha_1\alpha_2}\delta_{\beta_1\beta_2}}{\Nc}
 = \Qs^2\,\varGamma(x_1,x_2) \,.
\end{equation}
Consequently, the singlet part of the two-point function of Wilson
lines in the fundamental representation is
\begin{equation}
 \bigl\langle U(\bx_{1\perp})_{\beta_1\alpha_1}
 U^\ast(\bx_{2\perp})_{\beta_2\alpha_2} \bigr\rangle
 = \exp\bigl[ -\Qs^2\,\varGamma(x_1,x_2) \bigr] \,
 \delta_{\beta_1\beta_2}\,\delta_{\alpha_1\alpha_2}/\Nc \,,
\label{eq:two-point}
\end{equation}
which means that only the closed color dipole survives under the
average over random color distribution inside a dense gluon medium.

%%%   <UU> in the general representation   %%%

\subsection{$\langle U(\bx_{1\perp})_{\beta_1\alpha_1}
 U(\bx_{2\perp})_{\beta_2\alpha_2}\rangle$ in the general
 representation}

  It is possible to extend the previous argument formally.  We can
reach a more general expression in arbitrary representation.  The
direct product of two irreducible representations, $r_1$ and $r_2$,
makes one singlet when $r_2$ is the star representation to $r_1$, that
is, $r_1=r_2^*=r$, leading to the generators, $t_2^a=-t_1^{a*}$.  In
this general case we have
$\langle\alpha_1\alpha_2|s\rangle=\delta_{\alpha_1\alpha_2}/\sqrt{d_r}$,
where $d_r$ is the dimension of $r$ representation, which is given by
the Dynkin coefficients $m$ charactering the irreducible
representation $r$ as
\begin{equation}
 d_r(m) = \prod_{i<j}^{\Nc} \Biggl(1+\sum_{n=i}^{j-1}
  \frac{m_n}{j-i} \Biggr) \,.
\end{equation}
We can express $t^a_1 t_2^a$ using the Casimir invariant;
\begin{equation}
 t^a_1\,t_2^a = \frac{1}{2}\bigl[ (t^a_1+t^a_2)^2-(t_1^a)^2-(t_2^a)^2
  \bigr] = -\frac{1}{2}\bigl[ C_2(r_1)+C_2(r_2) \bigr] = -C_2(r) \,,
\end{equation}
leading to
\begin{equation}
 \langle s|V|s\rangle = \Qs^2\frac{2\Nc}{\Nc^2-1}\,C_2(r)\,
  \varGamma(x_1,x_2) \,.
\end{equation}
After all, the generalization of Eq.~(\ref{eq:two-point}) takes an
expression of
\begin{equation}
 \bigl\langle U(\bx_{1\perp})_{\beta_1\alpha_1}
  U(\bx_{2\perp})_{\beta_2\alpha_2} \bigr\rangle
  = \exp\biggl[
  -\Qs^2\,\frac{2\Nc}{\Nc^2-1}\,C_2(r)\,\varGamma(x_1,x_2)
  \biggr]\,\delta_{\beta_1\alpha_1}\,\delta_{\beta_2\alpha_2}/d_r \,.
\end{equation}

%%%   <UU...U> in the fundamental representation   %%%

\subsection{$\langle U(\bx_{1\perp})_{\alpha_1\beta_1}
 U(\bx_{2\perp})_{\alpha_2\beta_2}\cdots
 U(\bx_{\Nc\perp})_{\alpha_{\Nc}\beta_{\Nc}}\rangle$ in the
 fundamental representation}

  This expectation value is relevant to the scattering amplitude
between a baryon consisting of $\Nc$ valence quarks and dense gluon
matter inside a heavy hadron.  One can find an example of the baryon
expectation value in the context of the Odderon
physics~\cite{Jeon:2005cf,Hatta:2005as}, though the Gaussian average
can only describe the Pomeron part.

  The singlet state with one baryon is
$\langle\alpha_1\cdots\alpha_n|s\rangle
=\epsilon_{\alpha_1\alpha_2\cdots\alpha_{\Nc}}/\sqrt{\Nc!}$ where
$\epsilon_{\alpha_1\alpha_2\cdots\alpha_{\Nc}}$ is the antisymmetric
tensor with the definition $\epsilon_{12\cdots\Nc}=1$.
The matrix element $\langle s|{\tf}^a_i {\tf}^a_j |s\rangle$ does not
depend on the indices, $i$ and $j$, that is,
\begin{equation}
 \langle s|{\tf}^a_i\,{\tf}^a_j|s\rangle =
  -\frac{{\tf}^a_{\beta\alpha}{\tf}^a_{\alpha\beta}}{\Nc!}\,
  (\Nc-2)! = -\frac{\Nc+1}{2\Nc} \,,
\end{equation}
that yields
\begin{equation}
 \langle s|V|s\rangle = \Qs^2\,\frac{1}{\Nc-1}\sum_{i>j}\,
  \varGamma(\bx_{i\perp},\bx_{j\perp}) \,.
\end{equation}
The baryon expectation value is thus,
\begin{eqnarray}
 &\displaystyle \bigl\langle U(\bx_{1\perp})_{\beta_1\alpha_1}
  U(\bx_{2\perp})_{\beta_2\alpha_2}\cdots
  U(\bx_{\Nc\perp})_{\beta_{\Nc}\alpha_{\Nc}} \bigr\rangle
  = \qquad\qquad \nonumber\\
 &\displaystyle \qquad\qquad
  = \exp\biggl[ -\Qs^2\,\frac{1}{\Nc-1}\sum_{i>j}\,
  \varGamma(\bx_{i\perp},\bx_{j\perp}) \biggr]
  \frac{\epsilon_{\alpha_1\alpha_2\cdots\alpha_{\Nc}}
  \epsilon_{\beta_1\beta_2\cdots\beta_{\Nc}}}{\Nc!} \,.
\end{eqnarray}
When $\Nc$ gets large, the exponential factor is decreasing in
contrast to the meson scattering in Eq.~(\ref{eq:two-point}) that
stays unsuppressed for large $\Nc$.  This is a manifestation of the
fact that baryons would not live as they are in the large-$\Nc$ limit
but mesons would.

%%%   <UUU> in the adjoint representation   %%%

\subsection{$\langle\tilde{U}(\bx_{1\perp})_{\beta_1\alpha_1}
 \tilde{U}(\bx_{2\perp})_{\beta_2\alpha_2}
 \tilde{U}(\bx_{3\perp})_{\beta_3\alpha_3}\rangle$ in the adjoint
 representation}

  So far, the number of singlet is only one, and the next step we are
heading for is to treat the case with multiple singlets.  The simplest
and still non-trivial is the three gluon propagation, in which
$(\Nc^2-1)\otimes(\Nc^2-1)\otimes(\Nc^2-1)$ includes two singlets
composed of
$f_{abc}=-2\rmi\tr([\ta^a,\ta^b]\ta^c)$ and
$d_{abc}=2\tr(\{\ta^a,\ta^b\}\ta^c)$, where $\ta^a$'s represent the
$\SU(\Nc)$ generator in the adjoint representation, as
\begin{eqnarray}
 \langle\alpha_1\alpha_2\alpha_3|s_1\rangle
  &=& f_{\alpha_1\alpha_2\alpha_3}\frac{1}{\sqrt{(\Nc^2-1)\Nc}}\,,
 \nonumber\\
 \langle\alpha_1\alpha_2\alpha_3|s_2\rangle
  &=& d_{\alpha_1\alpha_2\alpha_3}\,
  \sqrt{\frac{\Nc}{(\Nc^2-4)(\Nc^2-1)}}
\end{eqnarray}
with a proper normalization.  Using the Jacobi identities,
$f_{abe}f_{cde}+f_{ade}f_{bce}+f_{ace}f_{dbe}=0$ and
$f_{abe}d_{cde}+f_{ade}d_{bce}+f_{ace}d_{dbe}=0$ we can calculate each
matrix element in singlet space as
\begin{eqnarray}
 \langle s_1|{\ta}^a_i\,{\ta}^a_j|s_1\rangle &=&
  -f_{\alpha_1\,a\,\beta_1} f_{\alpha_2\,a\,\beta_2}
  f_{\alpha_1\alpha_2\beta_3} f_{\beta_1\beta_2\beta_3}\,
  \frac{1}{\Nc(\Nc^2-1)} = -\frac{\Nc}{2} \,,\nonumber\\
 \langle s_1|{\ta}^a_i\,{\ta}^a_j|s_2\rangle &=&
  -f_{\alpha_1\,a\,\beta_1} f_{\alpha_2\,a\,\beta_2}
  f_{\alpha_1\alpha_2\beta_3} d_{\beta_1\beta_2\beta_3}\,
  \frac{1}{(\Nc^2-1)\sqrt{\Nc^2-4}} = 0 \,,\nonumber\\
 \langle s_2|{\ta}^a_i\,{\ta}^a_j|s_2\rangle &=&
  -f_{\alpha_1\,a\,\beta_1} f_{\alpha_2\,a\,\beta_2}
  d_{\alpha_1\alpha_2\beta_3} d_{\beta_1\beta_2\beta_3}\,
  \frac{\Nc}{(\Nc^2-4)(\Nc^2-1)} = -\frac{\Nc}{2} \,,
\end{eqnarray}
which are independent of the indices $i$ and $j$, and thus the matrix
structure is simply proportional to unity.  The projected $V$ is then,
\begin{equation}
 V = \Qs^2\,\frac{\Nc^2}{\Nc^2-1} \sum_{i>j}\,
  \varGamma(\bx_{i\perp},\bx_{j\perp})
  \begin{pmatrix} 1 & 0 \\ 0 & 1 \end{pmatrix} \,.
\end{equation}
At last, we get the result,
\begin{eqnarray}
 \bigl\langle\tilde{U}(\bx_{1\perp})_{\beta_1\alpha_1}
  \tilde{U}(\bx_{2\perp})_{\beta_2\alpha_2}
  \tilde{U}(\bx_{3\perp})_{\beta_3\alpha_3}
  \bigr\rangle &=& \frac{1}{\Nc(\Nc^2-1)}\, \exp\biggl[
  -\Qs^2\,\frac{\Nc^2}{\Nc^2-1}\sum_{i>j}\,
  \varGamma(\bx_{i\perp},\bx_{j\perp}) \biggr] \times \nonumber\\
 &&\times \biggl( f_{\alpha_1\alpha_2\alpha_3} f_{\beta_1\beta_2\beta_3}
  +\frac{\Nc^2}{\Nc^2 \!-\! 4} d_{\alpha_1\alpha_2\alpha_3}
  d_{\beta_1\beta_2\beta_3} \biggr) \,.
\end{eqnarray}
Roughly speaking, three gluons can make two kinds of color singlet
glueballs and they do not mix together due to different symmetry.  The
first and second terms in the curly parenthesis are contributions from
those glueball states respectively.

%%%   <UU^*UU^*> in the fundamental representation   %%%

\subsection{$\langle U(\bx_{1\perp})_{\beta_1\alpha_1}
 U^\ast(\bx_{2\perp})_{\beta_2\alpha_2} U(\bx_{3\perp})_{\beta_3\alpha_3}
 U^\ast(\bx_{4\perp})_{\beta_4\alpha_4}\rangle$ in the
 fundamental representation}

  The next non-trivial example is the four-point Wilson lines in the
fundamental representation.  This problem reduces to the irreducible
decomposition of $\Nc\otimes\Nc^\ast\otimes\Nc\otimes\Nc^\ast$, which
contains two normalized singlets;
\begin{eqnarray}
 \langle\alpha_1\alpha_2\alpha_3\alpha_4|s_1\rangle
  &=& \frac{1}{\Nc}\,\delta_{\alpha_1\alpha_2}\delta_{\alpha_3\alpha_4}
  \,,\nonumber\\
 \langle\alpha_1\alpha_2\alpha_3\alpha_4|s_2\rangle
  &=& \frac{1}{\sqrt{\Nc^2-1}} \biggl(\delta_{\alpha_1\alpha_4}
  \delta_{\alpha_2\alpha_3}-\frac{1}{\Nc}\,\delta_{\alpha_1\alpha_2}
  \delta_{\alpha_3\alpha_4} \biggr) \,,
\label{eq:fourpointsinglet}
\end{eqnarray}
leading to the projected matrix elements of
\begin{eqnarray}
 V &=& -\Qs^2\,\frac{2\Nc}{\Nc^2-1}\Bigl(-{\tf}^a_1\,{\tf}^{a\ast}_2\,
  \varGamma(\bx_{1\perp},\bx_{2\perp}) + {\tf}^a_1\,{\tf}^a_3\,
  \varGamma(\bx_{1\perp},\bx_{3\perp}) - {\tf}^a_1\,{\tf}^{a\ast}_4\,
  \varGamma(\bx_{1\perp},\bx_{4\perp}) -\nonumber\\
&& -{\tf}^{a\ast}_2\,{\tf}^a_3\,\varGamma(\bx_{2\perp},\bx_{3\perp})
 + {\tf}^{a\ast}_2\,{\tf}^{a\ast}_4\,\varGamma(\bx_{2\perp},\bx_{4\perp})
 - {\tf}^a_3\,{\tf}^{a\ast}_4\,\varGamma(\bx_{3\perp},\bx_{4\perp})
 \Bigr) \,,
\end{eqnarray}
given by
\begin{equation}
 \begin{pmatrix} V_{11} & V_{12} \\ V_{21} & V_{22} \end{pmatrix} =
  -\Qs^2\,\frac{2\Nc}{\Nc^2-1}
 \begin{pmatrix}
  \displaystyle -\frac{\Nc^2-1}{\Nc}\gamma &
  \displaystyle \frac{\sqrt{\Nc^2-1}}{\Nc} \bigl( \beta-\alpha \bigr) \\
  \displaystyle \frac{\sqrt{\Nc^2-1}}{\Nc} \bigl( \beta-\alpha \bigr) &
  \displaystyle \qquad
   \frac{1}{\Nc}\Bigl(\gamma-2\beta+(2-\Nc^2)\alpha\Bigr)
 \end{pmatrix} \,,
\end{equation}
where we defined with slight modification from
Ref.~\cite{Blaizot:2004wv}
\begin{eqnarray}
 2\alpha &=& \varGamma(\bx_{1\perp},\bx_{4\perp})
  +\varGamma(\bx_{2\perp},\bx_{3\perp}) \,,\nonumber\\
 2\beta &=& \varGamma(\bx_{1\perp},\bx_{3\perp})
  +\varGamma(\bx_{2\perp},\bx_{4\perp}) \,,\\
 2\gamma &=& \varGamma(\bx_{1\perp},\bx_{2\perp})
  +\varGamma(\bx_{3\perp},\bx_{4\perp}) \,,\nonumber
\end{eqnarray}
which are free from infrared singularity.  In order to calculate the
matrix element of $\exp(-V)$, we need to diagonalize $V$ whose
eigenvalues are given as the solution of the characteristic equation
$\det[\lambda-V]=0$, which gives us two eigenvalues,
\begin{equation}
 \lambda_{\pm} = \frac{1}{2}\bigl(\tr V \pm \varphi \bigr)
\end{equation}
with
\begin{equation}
 \varphi = \sqrt{(\tr V)^2-4\det V} =
  \Qs^2\frac{2\Nc^2}{\Nc^2-1}\sqrt{(\alpha-\gamma)^2
  +\frac{4}{\Nc^2}(\beta-\alpha)(\beta-\gamma)}
\end{equation}
in our notation.  The eigenstates are
\begin{equation}
 u_+ = \begin{pmatrix} \cos\theta\\ \sin\theta \end{pmatrix}\,,
\qquad
 u_- = \begin{pmatrix} -\sin\theta\\ \cos\theta \end{pmatrix}\,,
\end{equation}
with
\begin{equation}
 \tan\theta = \frac{2V_{12}}{V_{11}-V_{22}+\varphi} \,.
\end{equation}
Using the unitary matrix $T=(u_+\;u_-)$ we can diagonalize $\exp(-V)$
to have
\begin{eqnarray}
 \rme^{-V} &=& T \begin{pmatrix} \rme^{-\lambda_+} & 0 \\
  0 & \rme^{-\lambda_-} \end{pmatrix} T^{-1} \nonumber\\
 &=& \rme^{-\frac{1}{2}\tr V} T \begin{pmatrix}
  \rme^{-\frac{1}{2}\varphi} & 0 \\
  0 & \rme^{\frac{1}{2}\varphi} \end{pmatrix} T^{-1} \nonumber\\
 &=& \rme^{-\frac{1}{2}\tr V} \begin{pmatrix}
  \cosh\half\varphi-(\cos^2\theta-\sin^2\theta)
  \sinh\half\varphi  &  -2\cos\theta\sin\theta\sinh\half\varphi \\
  -2\cos\theta\sin\theta\sinh\half\varphi  &  \cosh\half\varphi
  +(\cos^2\theta-\sin^2\theta)\sinh\half\varphi \end{pmatrix} \nonumber\\
 &=& \rme^{-\frac{1}{2}\tr V} \Biggl[\cosh\half\varphi
  -\frac{\sinh\half\varphi}{\varphi}
  \begin{pmatrix} V_{11}-V_{22} & 2V_{12} \\ 2V_{12} & -V_{11}+V_{22}
  \end{pmatrix} \Biggr] \,.
\end{eqnarray}

  We can find one direct application of this example in the quark
production from the CGC background in the p-A
collision~\cite{Blaizot:2004wv}, in which the necessary quantity is
\begin{eqnarray}
&&\hspace{-1cm} \bigl\langle \tr\bigl\{ U(\bx_{1\perp})\,\tf^a\,
  U^\dagger(\bx_{2\perp})\,U(\bx_{3\perp})\,\tf^a\,
  U^\dagger(\bx_{4\perp})\bigr\}\bigr\rangle = \nonumber\\
&=& {\tf}^a_{\beta_1\beta_2}{\tf}^a_{\beta_3\beta_4}
  \delta_{\alpha_2\alpha_3}\delta_{\alpha_4\alpha_1}
  \bigl\langle U(\bx_{1\perp})_{\alpha_1\beta_1}
  U^\ast(\bx_{2\perp})_{\alpha_2\beta_2} U(\bx_{3\perp})_{\alpha_3\beta_3}
  U^\ast(\bx_{4\perp})_{\alpha_4\beta_4} \bigr\rangle \,.
\end{eqnarray}
The color structure in ${\tf}^a_{\beta_1\beta_2}{\tf}^a_{\beta_3\beta_4}
\delta_{\alpha_2\alpha_3}\delta_{\alpha_4\alpha_1}$ is to be expressed
in terms of the singlet basis of Eq.~(\ref{eq:fourpointsinglet}) as
follows;
\begin{eqnarray}
&& {\tf}^a_{\beta_1\beta_2}{\tf}^a_{\beta_3\beta_4}\,
  \delta_{\alpha_2\alpha_3}\,\delta_{\alpha_4\alpha_1}
 = \nonumber\\
&&\qquad =\frac{\sqrt{\Nc^2-1}}{2}
  \langle\beta_1\beta_2\beta_3\beta_4|s_2\rangle
  \Bigl(\sqrt{\Nc^2-1}\langle s_2|\alpha_1\alpha_2\alpha_3\alpha_4\rangle
  +\langle s_1|\alpha_1\alpha_2\alpha_3\alpha_4\rangle \Bigr) \,.
\end{eqnarray}
Thus we immediately conclude
\begin{eqnarray}
 &&\hspace{-1cm} \bigl\langle \tr\bigl\{U(\bx_{1\perp})\,\tf^a\,
  U^\dagger(\bx_{2\perp})\,U(\bx_{3\perp})\,\tf^a\,
  U^\dagger(\bx_{4\perp}) \bigr\}\bigr\rangle =\nonumber\\
 &=& \frac{\Nc^2-1}{2}\biggl( \langle s_2|\rme^{-V}|s_2\rangle
  +\frac{1}{\sqrt{\Nc^2-1}}\langle s_1|\rme^{-V}|s_2\rangle \biggr)
  \nonumber\\
 &=& \frac{\Nc^2-1}{2}\,\rme^{-\frac{1}{2}\tr V}
  \biggl[ \cosh\half\varphi - \frac{\sinh\half\varphi}{\varphi}
  \Bigl( V_{22}-V_{11}+\frac{2}{\sqrt{\Nc^2-1}}V_{12}\Bigr)\biggr]
  \nonumber\\
 &=& \frac{\Nc^2\!-\!1}{2}\rme^{-\Qs^2\frac{\Nc^2}{\Nc^2-1}
  \bigl[-(\alpha+\gamma)+\frac{2}{\Nc^2}(\alpha+\gamma-\beta)\bigr]}\!
  \biggl[ \cosh\!\half\varphi + \frac{\Qs^2}{\varphi}\,
  \frac{2\Nc^2}{\Nc^2\!-\!1}(\gamma\!-\!\alpha)\sinh\!\half\varphi
  \biggr] .
\label{eq:fourpoint}
\end{eqnarray}
In order to compare this result to Eq.~(86) in
Ref.~\cite{Blaizot:2004wv}, we note that
$\Qs^2=(\Nc^2-1)/(4\Nc)\mu_A^2$ and
$\varphi=\half\Nc\mu_A^2|\alpha-\gamma|\sqrt{\Delta}$ if written with
the notation of Ref.~\cite{Blaizot:2004wv}.  Then, our
result~(\ref{eq:fourpoint}) turns out to agree exactly with
Ref.~\cite{Blaizot:2004wv}.

%%%   <UUUU> in the adjoint representation   %%%

\subsection{$\langle\tilde{U}(\bx_{1\perp})_{\beta_1\alpha_1}
 \tilde{U}(\bx_{2\perp})_{\beta_2\alpha_2}
 \tilde{U}(\bx_{3\perp})_{\beta_3\alpha_3}
 \tilde{U}(\bx_{4\perp})_{\beta_4\alpha_4}\rangle$ in the adjoint
 representation}

  This quantity is, literally speaking, the four-gluon scattering
amplitude.  Also, in the calculation of two-gluon production from the
CGC background~\cite{Fukushima:2007}, we have to evaluate
$\tr\{\tilde{U}\,\ta^a\,\tilde{U}^\dagger\tilde{U}\,\ta^a\,\tilde{U}^\dagger\bigr\}$
in the adjoint representation.  There are eight
singlets~\cite{Dittner} given by combinations of the following bases;
\begin{eqnarray}
 &\delta_{\alpha_1 \alpha_2}\delta_{\alpha_3 \alpha_4}\,,\qquad
 \delta_{\alpha_1 \alpha_3}\delta_{\alpha_2 \alpha_4}\,,\qquad
 \delta_{\alpha_1 \alpha_4}\delta_{\alpha_2 \alpha_3}\,,\nonumber\\
 &d^{\alpha_1 \alpha_2 \gamma}d^{\alpha_3 \alpha_4 \gamma}\,,\qquad
 d^{\alpha_1 \alpha_3 \gamma}d^{\alpha_2 \alpha_4 \gamma}\,,\\
 &d^{\alpha_1 \alpha_2 \gamma}f^{\alpha_3 \alpha_4 \gamma}\,,\qquad
 d^{\alpha_1 \alpha_3 \gamma}f^{\alpha_2 \alpha_4 \gamma}\,,\qquad
 d^{\alpha_1 \alpha_4 \gamma}f^{\alpha_2 \alpha_3 \gamma}\,.\nonumber
\end{eqnarray}
These states are not orthogonal, and it is very hard to obtain the
eigenvalues and eigenstates of an $8\times8$ matrix $V$ to compute
$\rme^{-V}$.  In the next section, we will develop an approximation to
simplify the calculation.

%%%%%%%%%%   Large-Nc limit   %%%%%%%%%%

\section{Large-$\Nc$ limit}

  It is possible to construct any higher-dimensional representation
from the direct product of the fundamental (and anti-fundamental)
representation.  A well-known example is the adjoint representation
whose matrix representation can be given in terms of the fundamental
Wilson lines as
$2\mathrm{tr}[U(\bx_\perp)\,t^\beta U^\dagger(\bx_\perp)\,t^\alpha]$
where $\alpha$ and $\beta$ run from one to $\Nc^2-1$.  Here we shall
consider the arbitrary product of $U$'s in a general way in the
large-$\Nc$ limit.  As usual, then, among color singlets, the baryon
operator is dropped and only the meson-type operators remain
non-vanishing, as we have seen before.  That is, we shall focus on the
singlets out of
$\Nc\otimes\Nc^\ast\otimes\cdots\otimes\Nc\otimes\Nc^\ast$.

  The Gaussian average of our interest is
\begin{equation}
 \Bigl\langle \prod_{i=1}^n U(\bx_{i\perp})_{\beta_i\alpha_i}
  U^\ast(\by_{i\perp})_{\bar{\beta}_i\bar{\alpha}_i} \Bigr\rangle
  = \exp\bigl[-(H_0+V)\bigr]_{\beta_1\bar{\beta}_1\cdots
  \beta_n\bar{\beta}_n;\alpha_1\bar{\alpha}_1\cdots
  \alpha_n\bar{\alpha}_n} \,,
\end{equation}
where we can write
\begin{eqnarray}
 H_0 &=& \Qs^2\,\frac{2\Nc}{\Nc^2-1}\,L(0,0)\Biggl[\sum_{i=1}^n
  ({\tf}_i^a-{\tf}_{\bar{i}}^{a\ast})\Biggr]^2 \,,\\
 V &=& -\Qs^2\,\frac{2\Nc}{\Nc^2-1}\Biggl\{\sum_{i>j}^n
  \Bigl[ {\tf}^a_i\,{\tf}^a_j\,\varGamma(\bx_{i\perp},\bx_{j\perp})
  +{\tf}^{a\ast}_{\bar{i}}\,{\tf}^{a\ast}_{\bar{j}}\,
  \varGamma(\by_{i\perp},\by_{j\perp}) \Bigr] -\nonumber\\
 && -\sum_{i,j=1}^n {\tf}^a_i\,{\tf}^{a\ast}_{\bar{j}}\,
  \varGamma(\bx_{i\perp},\by_{j\perp}) \Biggr\} \,,
\end{eqnarray}
before taking the limit of large $\Nc$.  The second-order Casimir
operator in $H_0$ is
\begin{equation}
 \Biggl[\sum_{i=1}^n({\tf}^a_i-{\tf}^{a\ast}_{\bar{i}})\Biggr]^2 =
  \sum_{i=1}^n ({{\tf}_i^a}^2 + {{\tf}_{\bar{i}}^{a\ast}}^2)
  + 2\sum_{i>j}^n ({\tf}^a_i {\tf}^a_j + {\tf}^{a\ast}_{\bar{i}}
  {\tf}^{a\ast}_{\bar{j}}) - 2\sum_{i,j=1}^n {\tf}^a_i
  {\tf}^{a\ast}_{\bar{j}} \,.
\label{eq:casimirLargeN}
\end{equation}
The first term of R.H.S.\ in Eq.~(\ref{eq:casimirLargeN}) is just the
Casimir operator and thus proportional to a unit matrix,
\begin{equation}
 \sum_{i=1}^n ({{\tf}_i^a}^2+{{\tf}_{\bar{i}}^{a\ast}}^2) = 
  2n\frac{\Nc^2-1}{2\Nc} \;\to\; n\Nc \,,
\label{eq:first}
\end{equation}
in the large-$\Nc$ limit.  Next, let us check that the second term of
R.H.S.\ in Eq.~(\ref{eq:casimirLargeN}) leads to only
$\mathcal{O}(1)$ contributions.  A well-known formula reads
\begin{equation}
 {\tf}^a_{\beta_i\alpha_i}\,{\tf}^a_{\beta_j\alpha_j} = \frac{1}{2}
  \Bigl(\delta_{\beta_i\alpha_j}\delta_{\beta_j\alpha_i}
  -\frac{1}{\Nc}\delta_{\beta_i\alpha_i}\delta_{\beta_j\alpha_j}\Bigr)
 \;\to\; \frac{1}{2}\Bigl(
  \parbox{3cm}{\includegraphics[width=3cm]{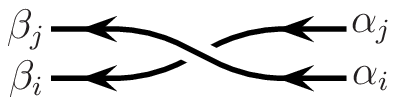}}\Bigr) \,.
\end{equation}
This cannot make a loop with any singlet state;  a loop formed with
$\alpha_i$-$\beta_j$ and $\alpha_j$-$\beta_i$ connected is not a
singlet with respect to $\alpha_i$'s or $\beta_j$'s which should be
accompanied by a huge suppression factor in the infrared sector.  This
is why the second term is only negligible in the large-$\Nc$ limit.
The last term of R.H.S.\ in Eq.~(\ref{eq:casimirLargeN}) has
$\mathcal{O}(\Nc)$ contributions as seen from
\begin{equation}
 -{\tf}^a_{\beta_i\alpha_i}\,{\tf}^{a\ast}_{\bar{\beta_j}\bar{\alpha_j}} =
  -\frac{1}{2}\Bigl(\delta_{\beta_i\bar{\beta_j}}
  \delta_{\alpha_i\bar{\alpha}_j}-\frac{1}{\Nc}
  \delta_{\beta_i\alpha_i}\delta_{\bar{\beta_j}\bar{\alpha_j}}\Bigr)
 \;\to\; -\frac{1}{2}\Bigl(
  \parbox{3cm}{\includegraphics[width=3cm]{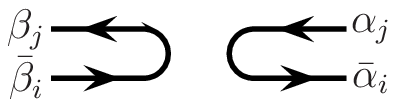}}\Bigr) \,,
\end{equation}
that can make a loop like
\begin{equation}
 -{\tf}^a_{\beta_i\alpha_i}\,{\tf}^{a\ast}_{\bar{\beta_j}\bar{\alpha_j}}
  \delta_{\alpha_j\bar{\alpha}_i} \;\to\; -\frac{1}{2}
  \Bigl(\parbox{3cm}{\includegraphics[width=3cm]{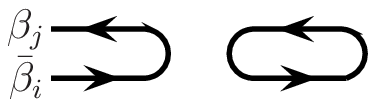}}\Bigr)
 = -\frac{\Nc}{2}\Bigl(
  \parbox{1.5cm}{\includegraphics[width=1.5cm]{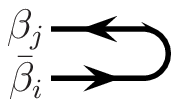}}\Bigr) \,.
\label{eq:third}
\end{equation}
From Eqs.~(\ref{eq:first}) and (\ref{eq:third}), we can find the color
bases on which $H_0$ vanishes.  There are $n!$ relevant singlets given
by all the permutations of
$\delta_{\alpha_{1}\bar{\alpha}_{1}}\cdots\delta_{\alpha_{n}\bar{\alpha}_{n}}$,
i.e.\ 
\begin{equation}
 \langle \alpha_1\cdots\alpha_n;\bar{\alpha}_1\cdots\bar{\alpha}_n
  |s_p\rangle = \frac{1}{\sqrt{\Nc^n}}\,
  \delta_{\alpha_1\bar{\alpha}_{p_1}} \cdots
  \delta_{\alpha_n\bar{\alpha}_{p_n}} \,,
\label{eq:singlet}
\end{equation}
where $(p_1,\dots,p_n)$ is the permutation of $(1,\dots,n)$ labeled
by $p$ which runs from one to $n!$.  Here we remark that we can relax
large $\Nc$ for Eq.~(\ref{eq:singlet}) being a singlet, though we
derived it in the large-$\Nc$ limit.

  It is easy to evaluate $V$ in this basis because the color matrix
structure of $V$ is quite similar to $H_0$ at large $\Nc$.  So that,
we have
\begin{equation}
 V|s_p\rangle \;\to\; \frac{2\Qs^2}{\Nc}\sum_{i,j=1}^n {\tf}^a_i\,
  {\tf}^{a\ast}_{\bar{j}}\,\varGamma(\bx_{i\perp},\by_{j\perp})
  |s_p\rangle \;\to\; \Qs^2\sum_{i=1}^n\varGamma(\bx_{i\perp},
  \by_{p_i\perp})|s_p\rangle \,.
\end{equation}
where we used Eq.~(\ref{eq:third}).  We finally arrive at the
expression of the Gaussian averaged Wilson loops in the large-$\Nc$
limit as follows;
\begin{equation}
 \Bigl\langle \prod_{i=1}^n U(\bx_{i\perp})_{\beta_i\alpha_i}
  U^\ast(\by_{i\perp})_{\bar{\beta}_i\bar{\alpha}_i} \Bigr\rangle
 \;\to\; \frac{1}{\Nc^n} \sum_{\{p\}=1}^{n!}\prod_{i=1}^n
  \delta_{\alpha_i\bar{\alpha}_{p_i}}\delta_{\beta_i\bar{\beta}_{p_i}}
  \exp\Biggl[-\Qs^2\sum_{j=1}^n \varGamma(\bx_{j\perp},\by_{p_j\perp})
  \Biggr] \,.
\label{eq:largeNc}
\end{equation}

  We shall apply our formula~(\ref{eq:largeNc}) for evaluation of the
expectation value of the dipole operator defined by
\begin{equation}
 D(\bx_\perp,\by_\perp) = \frac{1}{\Nc}\tr\bigl[ U(\bx_\perp)
  U^\dagger(\by_\perp) \bigr] = \frac{1}{\Nc}\delta_{\beta\bar{\beta}}\,
  \delta_{\alpha\bar{\alpha}}\, U(\bx_\perp)_{\beta\alpha}
  U^\ast(\by_\perp)_{\bar{\beta}\bar{\alpha}} \,.
\end{equation}
The scattering amplitude of the light projectile with $n$ color
dipoles is written as
\begin{equation}
 \Bigl\langle \prod_{i=1}^n D(\bx_{i\perp},\by_{i\perp})\Bigr\rangle =
  \exp\Biggl[ -\Qs^2\sum_{i=1}^n \varGamma(\bx_{i\perp},\by_{i\perp})
  \Biggr] \,.
\end{equation}
It should be mentioned that, when $\by_{i\perp}\to\bx_{i\perp}$, the
exponential factor becomes one with
$\varGamma(\bx_{i\perp},\by_{i\perp}\to\bx_{i\perp})\to0$, meaning
color transparency.

  We shall next compute the $n$-point Wilson line correlator in the
adjoint representation, which is as easy as
\begin{eqnarray}
 &&\hspace{-1cm}
  \Bigl\langle \prod_{i=1}^n \tilde{U}(\bx_{n\perp})_{b_i a_i}
  \Bigr\rangle = \nonumber\\
 &=& 2^n\prod_{i=1}^n 
  {\tf}^{b_i}_{\bar{\beta}_i\beta_i}
  {\tf}^{a_i}_{\alpha_i\bar{\alpha}_i}
  \bigl\langle U(\bx_{i\perp})_{\beta_i\alpha_i}
  U^\ast(\bx_{i\perp})_{\bar{\beta}_i\bar{\alpha}_i} \bigr\rangle
  \nonumber\\
 &=& 2^n\sum_{\{p\}}\prod_{i=1}^n 
  {\tf}^{b_i}_{\bar{\beta}_i\beta_i}\,
  {\tf}^{a_i}_{\alpha_i\bar{\alpha}_i}\,
  \frac{1}{\Nc^n}\, \delta_{\alpha_i\bar{\alpha}_{p_i}}
  \delta_{\beta_i\bar{\beta}_{p_i}} \exp\Biggl[ -\Qs^2\sum_{j=1}^n
  \varGamma(\bx_{j\perp},\bx_{p_j\perp})\Biggr] \nonumber\\
 &=& \frac{2^n}{\Nc^n} \sum_{\{p\}}\prod_{i=1}^n
  {\tf}^{a_i}_{\bar{\alpha}_{p_i}\bar{\alpha}_i}\,
  {\tf}^{b_i}_{\bar{\beta}_i\bar{\beta}_{p_i}}
  \exp\Biggl[ -\Qs^2\sum_{j=1}^n
  \varGamma(\bx_{j\perp},\bx_{p_j\perp})\Biggr] \,.
\end{eqnarray}

  Finally, we make a comment on the calculation of two-gluon
production from the CGC background.  In such a case, unfortunately,
the leading-$\Nc$ order of $\tr\{\tilde{U}\ta^a\tilde{U}^\dagger
\tilde{U}\ta^a\tilde{U}^\dagger\}$ is just vanishing and the
sub-leading order is necessary.  We will report details
elsewhere~\cite{Fukushima:2007}.

%%%%%%%%%%   Summary   %%%%%%%%%%

\section{Summary}

  We have derived the general formula to compute the correlation
function of Wilson lines in the random distribution of color source,
i.e.\ in the McLerran-Venugopalan model.  We emphasize that our
technique is to be applicable whenever the CGC weight function is
approximated as a Gaussian.  Our formula would be quite useful in
calculations not only of the scattering amplitude but also of the
particle production from the CGC background.  The correlation function
or the scattering amplitude is strongly suppressed if the color
non-singlet irreducible representation of the $n$-particle initial or
final state is involved.  Hence, we only have to consider the singlet
part of the color structure associated with the Wilson line product in
order to evaluate the correlation function.  After all, the problem of
evaluation of the correlation function is simply reduced to
diagonalization of a color matrix using singlet bases.  We have
explicitly written the two-point function down in the general
representation, the three-point function in the fundamental
representation corresponding to baryon's scattering off the CGC
background, in the adjoint representation as well,  and the four-point
function in the fundamental and anti-fundamental representation which
is related to the $q\bar{q}$ production from the CGC background.

  The larger number of Wilson lines, $n$, is involved in the Gaussian
average, the more difficult it is to find the singlets and to
diagonalize the color matrix.  As a matter of fact, the number of the
singlet states increases exponentially with increasing $n$, although
our method is powerful enough to implement also in numerical
computations in such an intricate case.  Instead of that, we took
advantage of simplifying the expressions in the large-$\Nc$ limit.  We
have derived the explicit formula with arbitrary number of Wilson
lines in the fundamental representation at large $\Nc$, from which, in
principle, any representation can be constructed.  For the
phenomenological application, we plan to address our calculation of
the two-gluon production from the CGC background in another
publication~\cite{Fukushima:2007} where we will discuss the
forward-backward rapidity correlation with respect to the hadron
multiplicity in the collision.

%%%%%%%%%%%%%%%%%%%%%%%%%%%%%%%%%%%%%%%%%%%%%%%%%%%%%%%%%%%%%%%%%%
\acknowledgments
  We thank Larry McLerran for giving us a tough problem, as a part of
which we initiated this work.  We also thank Raju Venugopalan for
discussions.  This research was supported in part by RIKEN BNL
Research Center and the U.S.\ Department of Energy under cooperative
research agreement \#DE-AC02-98CH10886.

%%%%%%%%%%%%%%%%%%%%%%%%%%%%%%%%%%%%%%%%%%%%%%%%%%%%%%%%%%%%%%%%%%


\begin{thebibliography}{99}

\bibitem{Nachtmann:1991ua}
  O.~Nachtmann, \ap{209}{1991}{436}.

\bibitem{Hufner:1990uv}
  J.~Hufner, C.~H.~Lewenkopf and M.~C.~Nemes, \npa{518}{1990}{297}.

\bibitem{Fujii:2002cj}
  H.~Fujii and T.~Matsui, \plb{545}{2002}{82}.

\bibitem{Fujii:2002vh}
  H.~Fujii, \npa{709}{2002}{236}.

\bibitem{McLerran:1993ni}
  L.~D.~McLerran and R.~Venugopalan, \prd{49}{1994}{2233};
  \ibid{49}{1994}{3352}; \ibid{50}{1994}{2225}.

\bibitem{Kovchegov:1997pc}
  Y.~V.~Kovchegov, \prd{54}{1996}{5463}; \ibid{55}{1997}{5445}.

\bibitem{Kovner:2001vi}
  A.~Kovner and U.~A.~Wiedemann, \prd{64}{2001}{114002}

\bibitem{Gelis:2001da}
  F.~Gelis and A.~Peshier,
  \npa{697}{2002}{879}.

\bibitem{Blaizot:2004wv}
  J.~P.~Blaizot, F.~Gelis and R.~Venugopalan,
  \npa{743}{2004}{57}.

\bibitem{Baier:2005dv}
  R.~Baier, A.~Kovner, M.~Nardi and U.~A.~Wiedemann,
  \prd{72}{2005}{094013}.

\bibitem{Fukushima:2007}
  K.~Fukushima and Y.~Hidaka, under completion.

\bibitem{Iancu:2002aq}
  E.~Iancu, K.~Itakura and L.~McLerran,
  \npa{724}{2003}{181}

\bibitem{Jeon:2004rk}
  S.~Jeon and R.~Venugopalan,
  \prd{70}{2004}{105012}.

\bibitem{Jeon:2005cf}
  S.~Jeon and R.~Venugopalan,
  \prd{71}{2005}{125003}.

\bibitem{Hatta:2005as}
  Y.~Hatta, E.~Iancu, K.~Itakura and L.~McLerran,
  \npa{760}{2005}{172}.

\bibitem{Dittner}
 P.~Dittner, \cmp{22}{1971}{238}.

\end{thebibliography}
\end{document}